\documentclass[10pt,a4paper]{article}

\usepackage[T1]{fontenc}
\usepackage{float}
\usepackage[utf8]{inputenc}
\usepackage{newtxtext,newtxmath}       
\usepackage{geometry}
\usepackage{graphicx}
\usepackage{subcaption}
\usepackage{amsmath}
\usepackage{upgreek}
\usepackage{parskip}
\usepackage{cite}
\usepackage[compact]{titlesec}
\usepackage[colorlinks=true,linkcolor=blue,citecolor=blue,urlcolor=blue]{hyperref}

\titlespacing{\section}{0pt}{6mm}{3mm}
\titlespacing{\subsection}{0pt}{6mm}{3mm}
\titlespacing{\subsubsection}{0pt}{6mm}{3mm}

\title{\bfseries Obstacle Avoidance of UAV in Dynamic Environments Using Direction and Velocity-Adaptive Artificial Potential Field}

\author{
    Nikita Vaibhav Pavle$^*$,
    Rakesh Kumar Sahoo$^\dagger$,
    Manoranjan Sinha$^\ddagger$\\[2mm]
    $^*$ Undergraduate Student, Department of Aerospace Engineering,\\
    Indian Institute of Technology Kharagpur, West Bengal, 721302, India\\
    $^\dagger$ PhD Research Scholar, Department of Aerospace Engineering,\\
    Indian Institute of Technology Kharagpur, West Bengal, 721302, India\\
    $^\ddagger$ Professor, Department of Aerospace Engineering,\\
    Indian Institute of Technology Kharagpur, West Bengal, 721302, India
}

\renewenvironment{abstract}{%
    \noindent\textbf{Abstract—}\normalsize
}{\par\vspace{1em}}

\date{}
\begin{document}
\maketitle

\begin{abstract}
The conventional Artificial Potential Field (APF) is fundamentally limited by the local minima issue and its inability to account for the kinematics of moving obstacles. This paper addresses the critical challenge of autonomous collision avoidance for Unmanned Aerial Vehicles (UAVs) operating in dynamic and cluttered airspace by proposing a novel Direction and Relative Velocity Weighted Artificial Potential Field (APF). In this approach, a bounded weighting function based on the relative direction and velocity of the UAV w.r.t the obstacle is introduced to dynamically scale the repulsive potential. This robust APF formulation is integrated within a Model Predictive Control (MPC) framework to generate collision-free trajectories while adhering to kinematic constraints. The simulation results demonstrate that the proposed method effectively resolves local minima and significantly enhances safety by allowing smooth, predictive avoidance maneuvers. The system ensures superior path integrity and reliable performance, confirming its viability for autonomous navigation in complex environments.
\end{abstract}

\section{Introduction}

The proliferation of Unmanned Aerial Vehicles (UAVs) marks a revolutionary stage in aerial technology, transitioning from specialized military uses to a broad spectrum of civilian and commercial applications. This expansion, spanning logistics, surveillance, infrastructure inspection, and even passenger transport, demonstrates the immense potential of autonomous flight. Achieving widespread, practical deployment of UAVs hinges entirely on the development of highly reliable, efficient, and sophisticated autonomous navigation systems. For UAVs to operate safely and practically in increasingly complex and cluttered airspace, overcoming the critical challenge of collision avoidance through robust real-time path planning is a fundamental prerequisite. This requires systems that can instantly perceive the environment and make smart decisions about movement.

Developing effective collision avoidance necessitates exploring a variety of path planning methodologies, which are typically categorized as global or local. Global planners, such as the Rapidly-exploring Random Tree (RRT)~\cite{yang2022rrt}, are known for their ability to find complete, collision-free paths in complex, static environments. However, these methods rely on a complete map of the environment and often require extensive computational resources. In stark contrast, local, reactive methods like the classic Artificial Potential Field (APF) technique are widely used due to their inherent reactive nature and remarkably low computational cost. APF operates by modeling the environment as a potential energy landscape, where the destination provides an attractive potential and obstacles generate repulsive potentials. Despite its simplicity, the conventional APF approach is severely hampered by the local minima issue, a condition where the attractive and repulsive forces reach an equilibrium point, causing the UAV to become prematurely trapped. While various complex solutions---including modifications~\cite{sahoo2025satnav,uppal2025eapf} like the black-hole~\cite{yao2020path} or vortex potential fields~\cite{nasuha2022vortex}, or integration with global planners like RRT~\cite{zhang2024rrtapf}---have been proposed, these often introduce significant computational overhead.

To address the fundamental limitations of the conventional APF---particularly its lack of consideration for dynamic threat---in highly dynamic environments, we introduce a novel Direction and Relative Velocity Weighted APF approach. This method significantly enhances the safety and efficiency of collision avoidance by modifying the standard repulsive force using a bounded weighting function. This function is designed to dynamically scale the base repulsive potential, based on two critical, real-time factors: the obstacle's directional threat relative to the UAV's motion and the instantaneous relative velocity threat between the UAV and the obstacle (captured by the velocity term. To address the fundamental limitations of the conventional APF—particularly its lack of consideration for dynamic threat—in highly dynamic environments, we introduce a novel Direction and Relative Velocity Weighted APF approach. This method significantly enhances the safety and efficiency of collision avoidance by modifying the standard repulsive force using a bounded weighting function. This function is designed to dynamically scale the base repulsive potential, based on two critical, real-time factors: the obstacle's directional threat relative to the UAV's motion and the instantaneous relative velocity threat between the UAV and the obstacle. 

This paper is organized as follows, detailing the proposed autonomous navigation system, its implementation, and performance evaluation. Section~\ref{sec:model} introduces the application of conventional Artificial Potential Field (APF) for Unmanned Aerial Vehicles (UAVs) in a dynamic environment and its limitations, such as the local minima issue. The translational and rotational kinematics and dynamics of UAVs are presented, followed by the formulation of Model Predictive Control (MPC) to compute an optimal safe position to avoid collision with an obstacle. Section~\ref{sec:apf} outlines the conventional APF and the local minima issue prevalent in the algorithm. Consequently, we introduce our novel direction and velocity adaptive APF, ensuring that the proposed repulsive potential function is bounded and enhances collision-avoidance manoeuvres. Simulation results for two different test scenarios---static local minima tests and moving obstacles---are presented in Section~\ref{sec:sim}. Finally, we conclude the work by summarizing the contribution of direction and velocity adaptive APF and suggesting avenues for future research.

\section{System Modeling and Control Architecture}
\label{sec:model}

Based on the physical structure of the quadcopter used for simulation~\cite{zhou2024cooperative}, we establish the dynamic and kinematic model in the Earth-fixed inertial frame ($W$) and the body-fixed frame ($B$), as detailed in the source literature.

\subsection{UAV Dynamic and Kinematic Modeling}

The position vector of the UAV is denoted by $p=[x,y,z]^T \in \mathbb{R}^3$.

\subsubsection{Linear Motion Dynamics}

The linear motion is governed by Newton's second law, relating the total external forces to the centre-of-mass acceleration. The total thrust generated by the four rotors is $F_{T} = \sum F_{i}$, where $F_{i}$ is the lift force of rotor $i$. The equations of motion in the Earth-fixed frame are:
\begin{equation}
m\ddot{p} = \begin{pmatrix} 0 \\ 0 \\ G \end{pmatrix} +
\mathbf{R}_{\text{B}}^{\text{W}} \begin{pmatrix} 0 \\ 0 \\ -F_T \end{pmatrix}
\label{eq:linear_motion}
\end{equation}
where $m$ is the mass of the drone, $G=mg$ is the gravitational force, and $\mathbf{R}_{\text{B}}^{\text{W}}$ is the rotation matrix transforming forces from the body frame ($B$) to the Earth-fixed frame ($W$), given by:
\begin{equation}
\mathbf{R}_{\text{B}}^{\text{W}} =
\begin{pmatrix}
c(\theta)c(\psi) & c(\phi)s(\psi) + s(\phi)s(\theta)c(\psi) & s(\phi)s(\psi) - c(\phi)s(\theta)c(\psi) \\
c(\theta)s(\psi) & c(\phi)c(\psi) - s(\phi)s(\theta)s(\psi) & s(\phi)c(\psi) + c(\phi)s(\theta)s(\psi) \\
-s(\theta) & c(\theta)s(\phi) & c(\theta)c(\phi)
\end{pmatrix}^T
\label{eq:rotation_matrix}
\end{equation}
where $c(\cdot)$ and $s(\cdot)$ represent $\cos(\cdot)$ and $\sin(\cdot)$ respectively, and $\phi, \theta, \psi$ are the roll, pitch, and yaw angles (Euler angles).

\subsubsection{Angular Motion Dynamics}

The angular motion is described by Euler's equation, where the total torque $\boldsymbol{\tau}$ is composed of the torques generated by rotor lift and the torsional drag moments. The body angular velocity vector is $\boldsymbol{\omega} = [p, q, r]^T$ around the $X_b, Y_b, Z_b$ axes, respectively:
\begin{equation}
\mathbf{J} \dot{\boldsymbol{\omega}} =
\begin{pmatrix} L(F_2 - F_4) \\ L(F_1 - F_3) \\ M_2 + M_4 - M_1 - M_3 \end{pmatrix}
- \boldsymbol{\omega} \times \mathbf{J} \boldsymbol{\omega}
\label{eq:angular_motion}
\end{equation}
where $L$ is the distance from the rotor centre to the centre of mass, $\mathbf{J}$ is the inertia matrix (or tensor), and $M_i$ is the torsional moment of rotor $i$.

Since the quadcopter control loop adopts an inner/outer loop architecture, the fast inner loop handles attitude control. For the outer loop, which focuses on position and velocity, the system is simplified as a second-order integrator (double integrator) model, where the control input is the desired acceleration $\mathbf{a}$.

\subsubsection{Continuous-Time Model}

For the UAV, the simplified system is:
\begin{equation}
\begin{cases}
\dot{p}(t) = v(t) \\
\dot{v}(t) = \mathbf{a}(t)
\end{cases}
\label{eq:simplified_cont}
\end{equation}
where $p=[x,y,z]^T \in \mathbb{R}^3$ is the position, $v=[v_x,v_y,v_z]^T \in \mathbb{R}^3$ is the velocity, and $\mathbf{a}=[a_x,a_y,a_z]^T \in \mathbb{R}^3$ is the acceleration control input.

This is represented in state-space form as:
\begin{equation}
\begin{cases}
\dot{x}(t) = A x(t) + B \mathbf{a}(t) \\
y(t) = C x(t)
\end{cases}
\label{eq:state_space_cont}
\end{equation}
with the state vector $x = [x, y, z, v_x, v_y, v_z]^T$ and input $\mathbf{a}=[a_x,a_y,a_z]^T$. The matrices are:
\[
A=\begin{pmatrix} 0 & I_3 \\ 0 & 0 \end{pmatrix}, \quad
B=\begin{pmatrix} 0 \\ I_3 \end{pmatrix}, \quad
C=\begin{pmatrix} I_3 & 0 \\ 0 & I_3 \end{pmatrix}.
\]

\subsubsection{Discrete-Time Model}

To apply MPC, the continuous-time model is discretized with a time interval $\Delta t$:
\begin{equation}
x(k+1) = A x(k) + B \mathbf{a}(k)
\label{eq:discrete_model}
\end{equation}
where the discrete system matrices $A$ and $B$ are approximated as:
\[
A=\begin{pmatrix} I_3 & I_3 \cdot \Delta t \\ 0 & I_3 \end{pmatrix}, \quad
B=\begin{pmatrix} \frac{1}{2}\Delta t^2 \cdot I_3 \\ \Delta t \cdot I_3 \end{pmatrix}.
\]

\subsection{Model Predictive Control (MPC) Formulation}

The MPC approach utilizes a prediction horizon of length $N$ to minimize a cost function $J_k$ by solving a constrained optimization problem for the optimal control sequence $\mathbf{U}_k$.

\subsubsection{State and Input Sequences}

The predicted states $x(k+i|k)$ and control inputs $\mathbf{a}(k+i|k)$ over the prediction horizon are combined into augmented vectors:
\begin{equation}
\mathbf{X}_k = M x(k) + C \mathbf{U}_k
\label{eq:combined_state}
\end{equation}
where $\mathbf{X}_k \in \mathbb{R}^{6(N+1)}$ is the combined state variable, $\mathbf{U}_k \in \mathbb{R}^{3N}$ is the combined input variable, and $M$ and $C$ are the prediction matrices derived from the discrete-time model (Eq.~\ref{eq:discrete_model}).

\subsubsection{Cost Function and Optimization}

The cost function $J_k$ is defined to minimize the tracking error (difference between predicted state $\mathbf{X}_k$ and the reference path $X_{\text{ref}}$ from APF) and the control effort $\mathbf{U}_k$:
\begin{equation}
J_k = [\mathbf{X}_k - X_{\text{ref}}]^T \mathbf{Q}_{\text{aug}} [\mathbf{X}_k - X_{\text{ref}}]
+ \mathbf{U}_k^T \mathbf{R}_{\text{aug}} \mathbf{U}_k
\label{eq:cost_function_matrix}
\end{equation}
where $X_{\text{ref}}$ contains the reference states (position and zero velocity) derived from the APF-generated prediction points $P^{\text{pred}}$. $\mathbf{Q}_{\text{aug}}$ and $\mathbf{R}_{\text{aug}}$ are augmented weighting matrices composed of the state weight $Q$, terminal state weight $F$, and control input weight $R$.

Substituting the state prediction (Eq.~\ref{eq:combined_state}) into the cost function yields the quadratic form:
\begin{equation}
J_k = G + 2 E \mathbf{U}_k + \mathbf{U}_k^T H \mathbf{U}_k
\label{eq:quadratic_cost}
\end{equation}
where $H = C^T \mathbf{Q}_{\text{aug}} C + \mathbf{R}_{\text{aug}}$ is the Hessian matrix, and $E$ and $G$ are terms dependent on the initial state $x(k)$ and the reference trajectory $X_{\text{ref}}$.

The Model Predictive Control problem is then formulated as a quadratic programming (QP) problem to find the optimal input sequence $\mathbf{U}_k$ subject to operational constraints (e.g., maximum speed, maximum acceleration):
\begin{equation}
\begin{split}
\min_{\mathbf{U}_k} \quad &\frac{1}{2} \mathbf{U}_k^T H \mathbf{U}_k + E^T \mathbf{U}_k \\
\text{s.t.} \quad & A_{\text{in}} \mathbf{U}_k \le b_{\text{in}}, \\
& A_{\text{eq}} \mathbf{U}_k = b_{\text{eq}}, \\
& lb \le \mathbf{U}_k \le ub.
\end{split}
\label{eq:qp_problem}
\end{equation}
The constraints $A_{\text{in}}, b_{\text{in}}, A_{\text{eq}}, b_{\text{eq}}, lb, ub$ encode limits on velocity, acceleration, and other state/control bounds over the prediction horizon.

\section{Direction and Velocity Weighted Artificial Potential Field}
\label{sec:apf}

\subsection{Classical Repulsive Potential Field}

Consider a quadrotor at position $\mathbf{p}$ and an obstacle at position $\mathbf{p}_o$.
Let $d = \|\mathbf{p}-\mathbf{p}_o\|$ be the distance to the obstacle and $d_s$ the
sphere of influence of the obstacle. The standard repulsive potential is
defined as
\begin{equation}
    U_{\mathrm{rep}}(d) =
    \begin{cases}
        \dfrac{1}{2}k_{\mathrm{rep}}\left(\dfrac{1}{d} - \dfrac{1}{d_s}\right)^2, & d \le d_s, \\[0.6em]
        0, & d > d_s,
    \end{cases}
    \label{eq:Urep}
\end{equation}
where $k_{\mathrm{rep}} > 0$ is a design gain.
The resulting repulsive force is obtained from the gradient of the potential:
\begin{equation}
    \mathbf{F}_{\mathrm{rep}} = - \nabla U_{\mathrm{rep}}(d).
\end{equation}

\subsection{Direction–Weighted APF}

The classical artificial potential field does not distinguish
whether the obstacle lies in front of or behind the UAV.
To bias the repulsive effect toward obstacles in the direction
of motion, we scale the repulsive potential with a
direction–dependent weight:
\begin{equation}
    U_{\mathrm{rep}}^{\ast}(d,\theta) = \omega(\theta)\, U_{\mathrm{rep}}(d),
\end{equation}
where $\theta$ is the angle between the UAV velocity vector
$\mathbf{v}$ and the relative position vector
$\mathbf{r}_e = \mathbf{p} - \mathbf{p}_o$, and
$\omega(\theta)$ is a direction–weighting function.

A convenient choice is
\begin{equation}
    \omega(\theta) = 1 + \gamma \frac{1+\cos\theta}{2}, 
    \qquad \gamma \ge 0,
    \label{eq:omega-theta}
\end{equation}
which satisfies
\begin{equation}
    1 \le \omega(\theta) \le 1+\gamma.
\end{equation}
Thus, $\omega(\theta)$ is maximal when $\theta = 0$ (obstacle directly in
front along the direction of motion) and minimal when $\theta = \pi$
(obstacle directly behind). The corresponding repulsive force is
\begin{equation}
    \mathbf{F}_{\mathrm{rep}}^{\ast} = - \nabla U_{\mathrm{rep}}^{\ast}(d,\theta).
\end{equation}

\subsection{Direction and Relative–Velocity Weighted APF}

A drawback of the purely direction–weighted APF is that it may
under–react to fast moving obstacles approaching from behind.
To address this, we further scale the potential with a term that
depends on the relative velocity between the UAV and the obstacle.

Define
\begin{equation}
    \mathbf{v}_e = \mathbf{v} - \mathbf{v}_o,
    \qquad
    \mathbf{r}_e = \mathbf{p} - \mathbf{p}_o,
    \qquad
    \hat{\mathbf{r}}_e = \frac{\mathbf{r}_e}{\|\mathbf{r}_e\|},
\end{equation}
and let
\begin{equation}
    G = \mathbf{v}_e \cdot \hat{\mathbf{r}}_e,
\end{equation}
which encodes both the magnitude of the relative velocity and whether
the obstacle is moving toward or away from the UAV.

The combined direction and relative–velocity weight is chosen as
\begin{equation}
    \omega(\theta,\mathbf{v}_e) =
    \left[1 + \gamma \frac{1+\cos\theta}{2}\right]
    \left( 2 + k \tanh(G) \right),
    \qquad \gamma \ge 0,\; k \ge 0,
    \label{eq:omega-theta-ve}
\end{equation}
so that the modified repulsive potential becomes
\begin{equation}
    U_{\mathrm{rep}}^{\ast}(d,\theta,\mathbf{v}_e)
    = \omega(\theta,\mathbf{v}_e)\, U_{\mathrm{rep}}(d).
\end{equation}

Since $\cos\theta \in [-1,1]$ and $\tanh(G) \in [-1,1]$, the weight is
bounded as
\begin{equation}
    1 \;\le\; \omega(\theta,\mathbf{v}_e)
    \;\le\; (1+\gamma)(2+k),
\end{equation}
with larger values obtained when the obstacle is located in front
of the UAV and moving toward it with high relative speed. The final
repulsive force used for collision avoidance is
\begin{equation}
    \mathbf{F}_{\mathrm{rep}}^{\ast}
    = - \nabla U_{\mathrm{rep}}^{\ast}(d,\theta,\mathbf{v}_e).
\end{equation}

\section{Simulation and Performance Analysis}
\label{sec:sim}

The proposed Direction and Relative Velocity Weighted APF approach is evaluated against the conventional Basic APF method across two distinct test scenarios: a static environment designed to induce local minima, and a general path-following test in a cluttered environment. Both methods use an identical Model Predictive Control (MPC) framework for trajectory tracking.

\subsection{Test Scenario 1: Static Local Minima Resolution}

This scenario tests the fundamental ability of the weighted APF to overcome traditional APF pitfalls using a specific, static obstacle arrangement.

\begin{figure}[h!]
    \centering
    \includegraphics[width=0.48\textwidth,height=7cm]{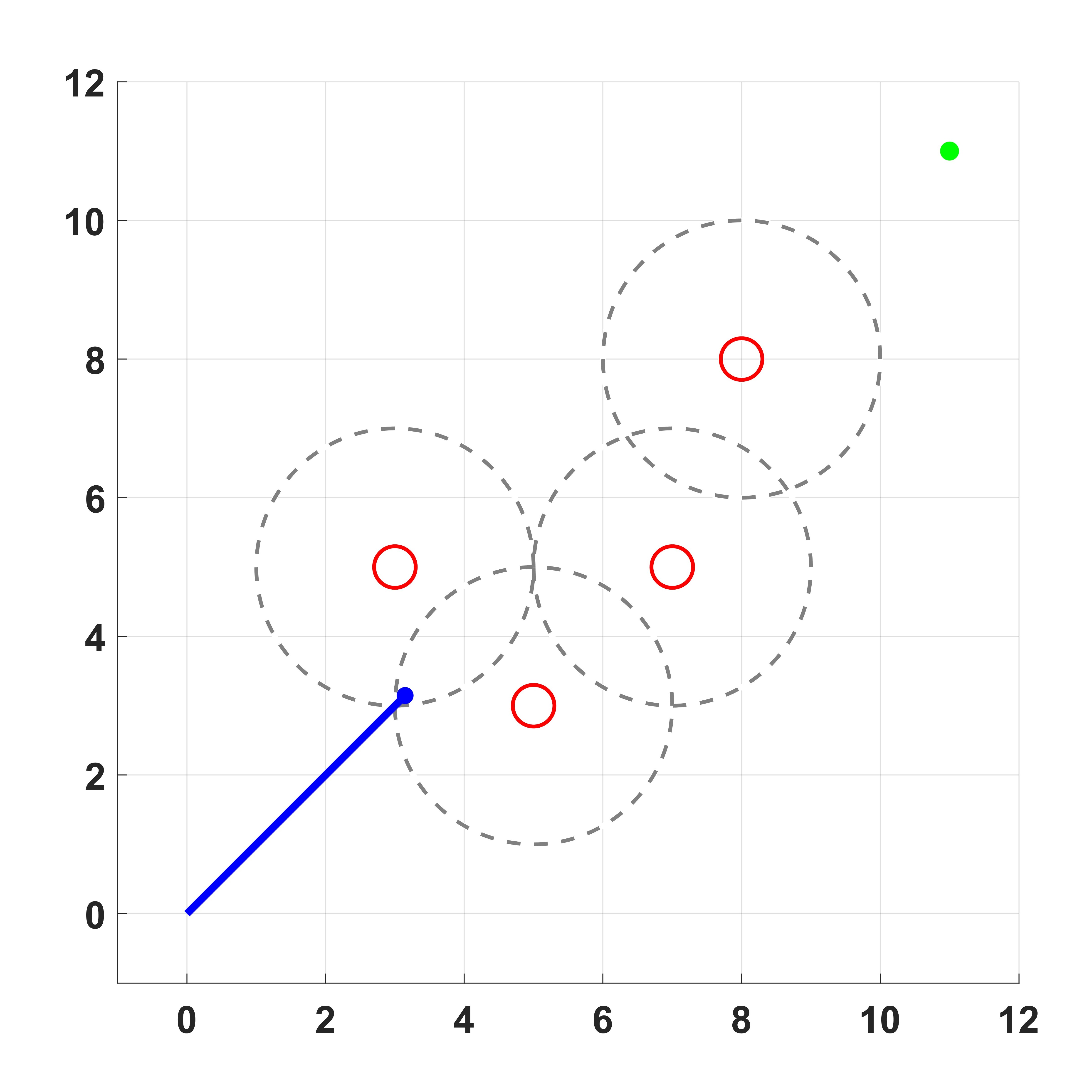}
    \hfill
    \includegraphics[width=0.48\textwidth,height=7cm]{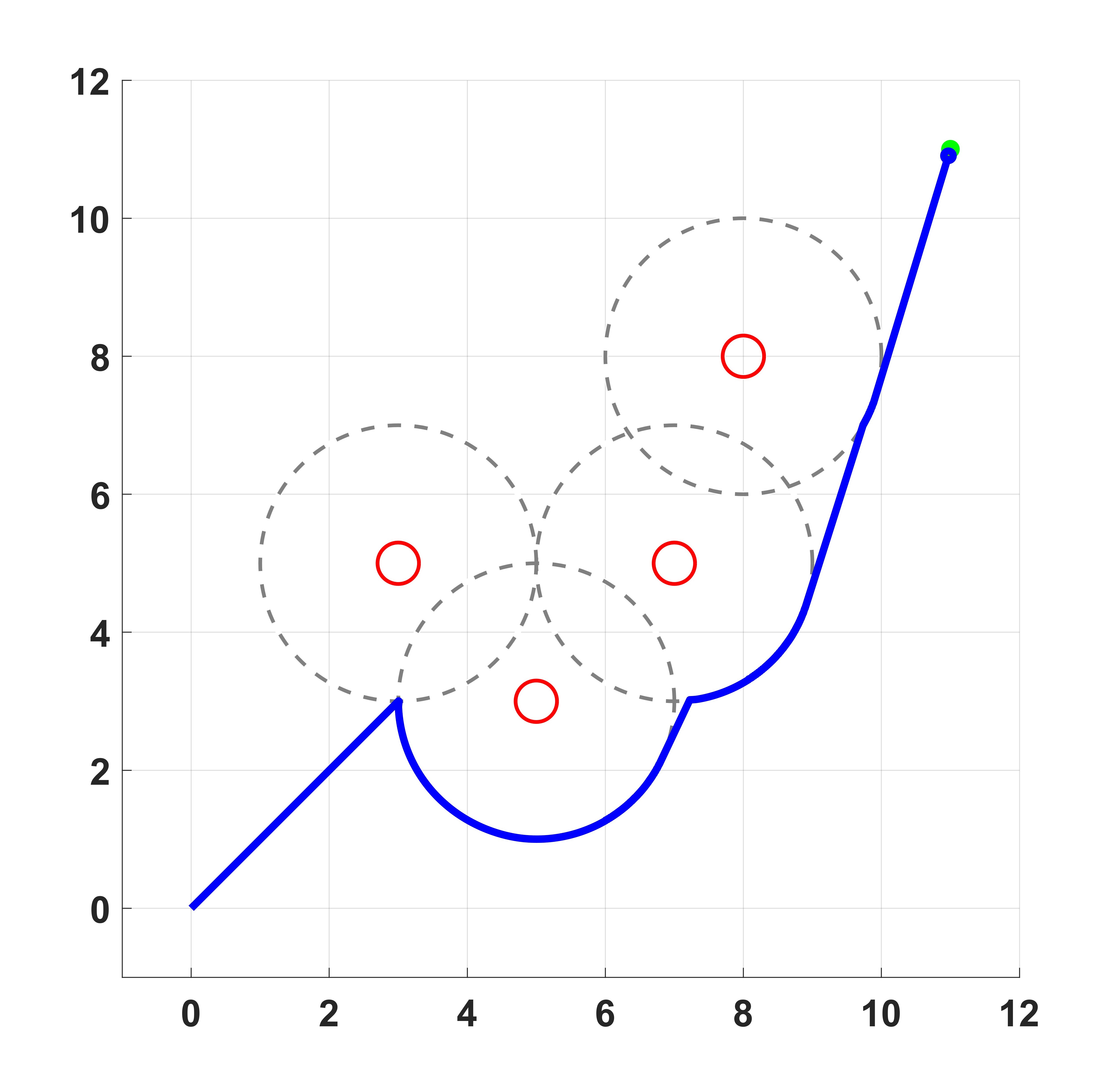}
    \caption{Comparison of local minima avoidance. (Left) Basic APF trajectory stalls/hesitates near the obstacle cluster (local minimum). (Right) Velocity-weighted APF successfully navigates around the cluster.}
    \label{fig:static_minima}
\end{figure}

\begin{itemize}
    \item \textbf{Basic APF performance (Figure~\ref{fig:static_minima}, left):} The trajectory clearly shows the UAV stalling when the attractive force to the goal (green dot) is balanced by the repulsive forces from the clustered obstacles (red circles). This confirms the presence of the classic local minima issue.
    \item \textbf{Velocity-weighted APF performance (Figure~\ref{fig:static_minima}, right):} In contrast, the velocity-weighted APF successfully navigates this exact configuration. By dynamically scaling the repulsive potential based on the vehicle's motion, the weighted function prevents the complete neutralization of forces, allowing the UAV to find a clear, smooth path around the obstacle cluster.
\end{itemize}

\subsection{Test Scenario 2: General Path Following and Efficiency}

This scenario uses a different arrangement of obstacles to evaluate the general path efficiency, safety margin, and control effort over a longer time horizon ($t=0$~s to $t=300$~s).

\subsubsection{Trajectory Comparison Over Time}

The time-stamped trajectory plots show the path taken by both methods in this general test scenario at $0.25, 0.5, 0.75$ and $1$ of $t$, where $t$ is the time taken to reach the goal.

\begin{figure}[h!]
    \centering
    \begin{subfigure}[b]{0.8\textwidth}
        \centering
        \includegraphics[width=\linewidth]{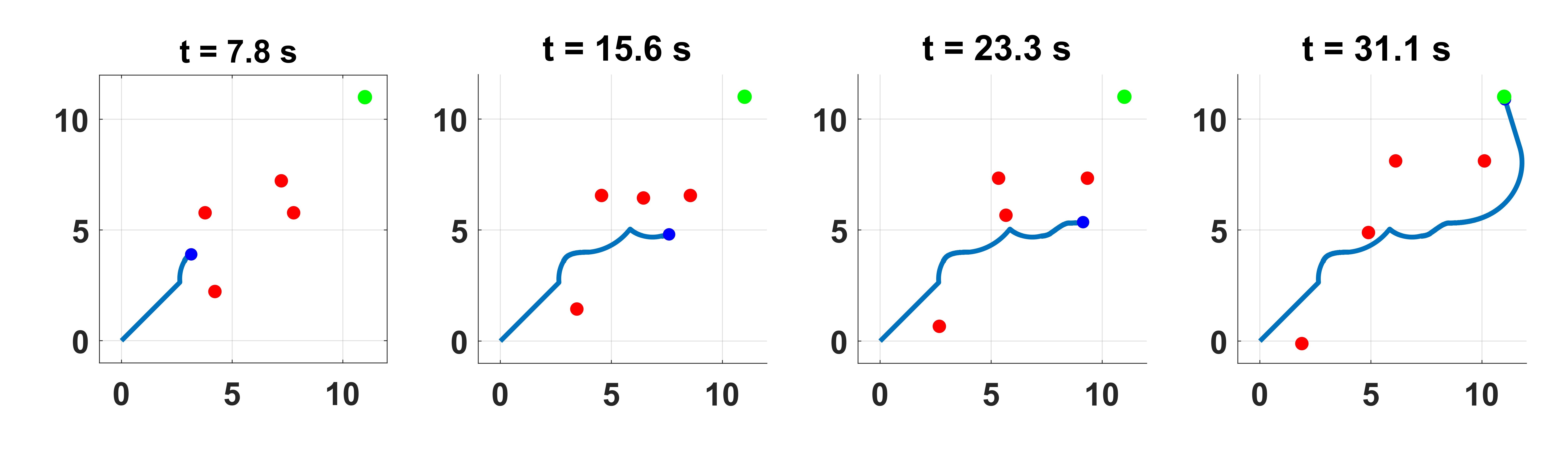}
        \caption{Timestamp for Basic APF.}
        \label{fig:uav_cases}
    \end{subfigure}
    
    \begin{subfigure}[b]{0.8\textwidth}
        \centering
        \includegraphics[width=\linewidth]{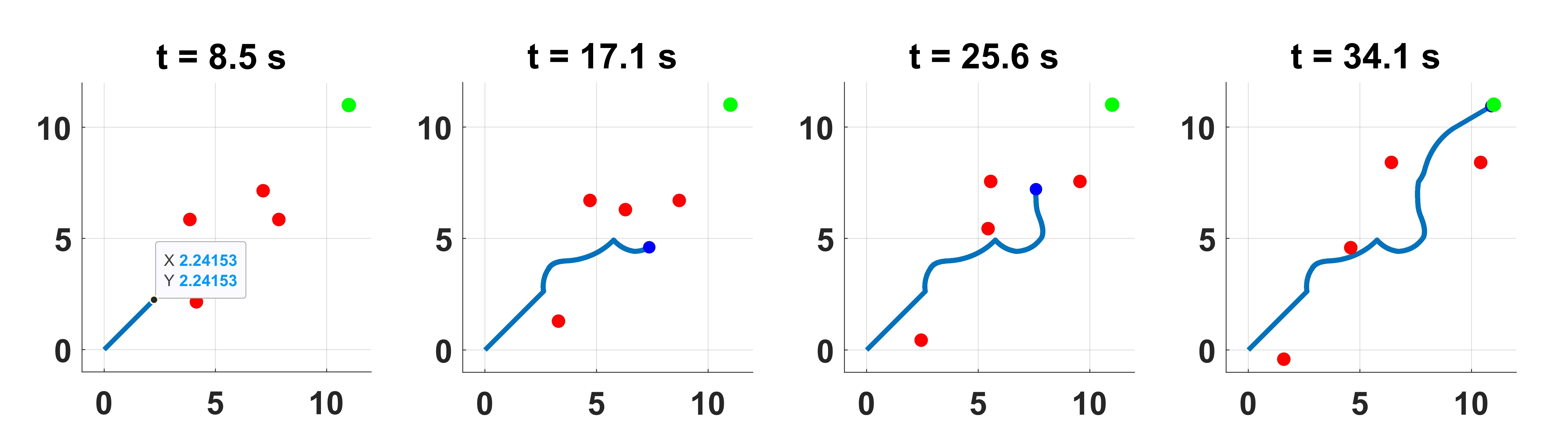}
        \caption{Timestamp for velocity-weighted APF.}
        \label{fig:uav_trajectory}
    \end{subfigure}
    \caption{Comparison between the path of Basic APF and velocity-based APF.}
    \label{fig:timestamp_trajectories}
\end{figure}

\begin{itemize}
    \item The time-stamped paths (Figure~\ref{fig:timestamp_trajectories}) confirm that both methods are capable of navigating the environment, but the velocity-weighted APF exhibits smoother overall curves, reflecting its more stable potential landscape.
\end{itemize}

\subsubsection{Obstacle Clearance and Safety Margin}

The distance-to-obstacle plots quantify the safety performance of the two algorithms.

\begin{figure}[h!]
    \centering
    \includegraphics[width=0.48\textwidth]{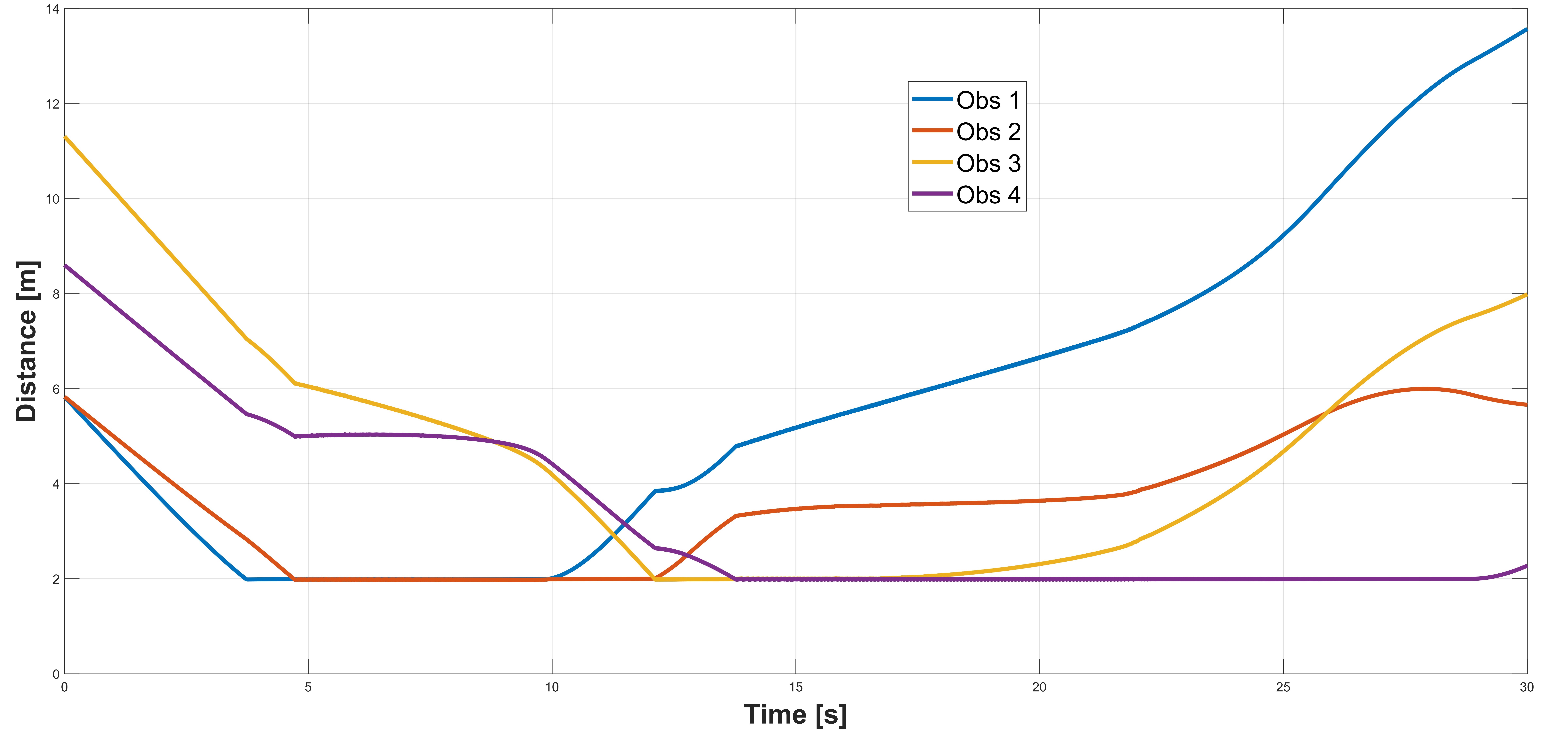}
    \hfill
    \includegraphics[width=0.48\textwidth]{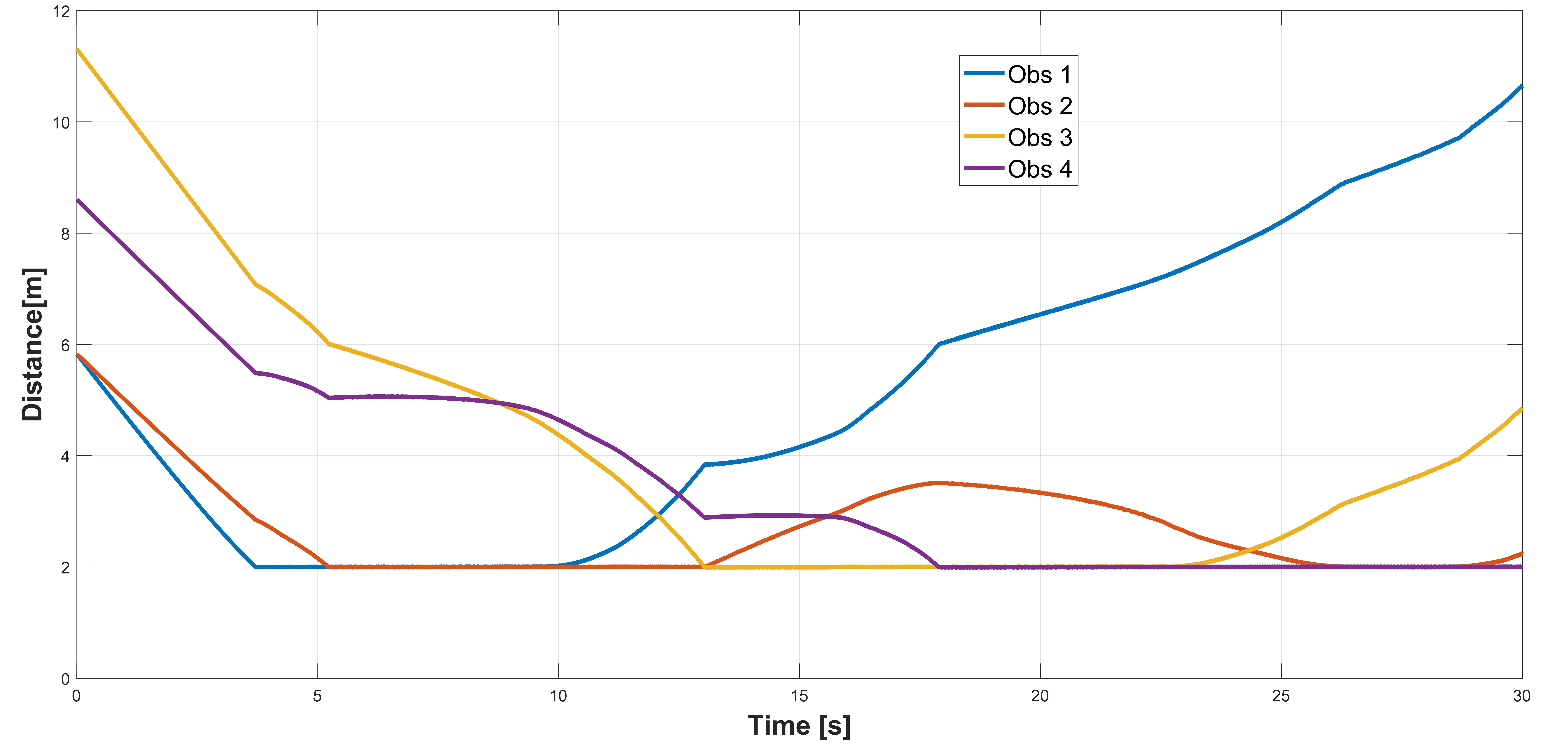}
    \caption{Distance to obstacles over time. (Left) Basic APF. (Right) Velocity-weighted APF.}
    \label{fig:obs_distance_comparison}
\end{figure}

\begin{itemize}
    \item \textbf{Basic APF (Figure~\ref{fig:obs_distance_comparison}, left):} This method exhibits several instances where the distance to obstacles (Obs 1 and Obs 4) rapidly drops to the minimum safe clearance limit ($\approx 2$~m). The sharp drop and prolonged time spent at minimum distance indicate a reactive, less predictive reliance on static distance.
    \item \textbf{Velocity-weighted APF (Figure~\ref{fig:obs_distance_comparison}, right):} This method achieves a generally smoother approach and recession from obstacles. The inclusion of the relative velocity term allows the algorithm to initiate avoidance manoeuvres earlier and more gradually when a high-speed collision threat is predicted, enhancing safety modulation.
\end{itemize}

\subsubsection{Goal Convergence and Path Efficiency}

The distance-to-goal plots confirm successful navigation and allow for a comparison of overall efficiency.

\begin{figure}[h!]
    \centering
    \includegraphics[width=0.48\textwidth]{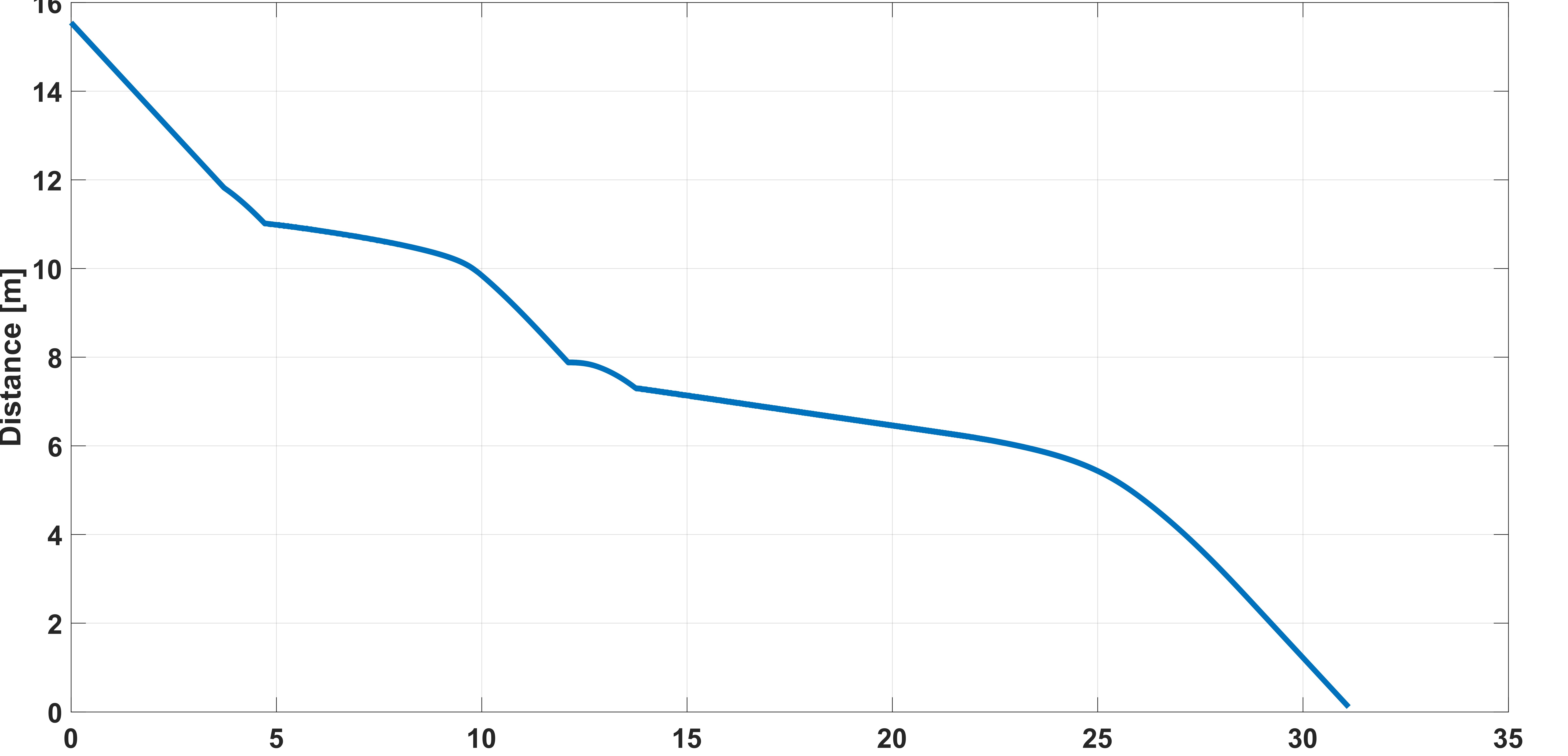}
    \hfill
    \includegraphics[width=0.48\textwidth]{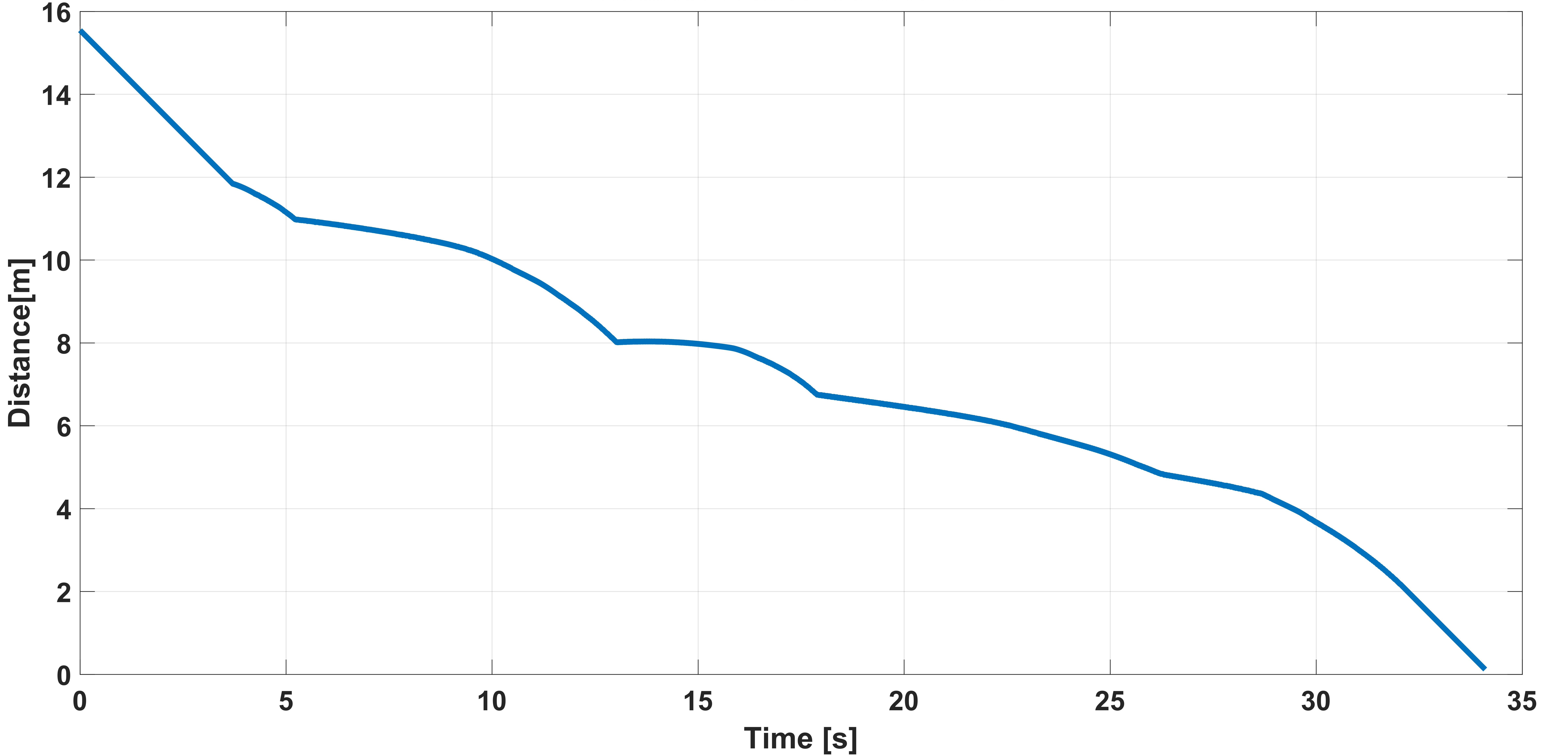}
    \caption{Distance to goal over time. (Left) Basic APF. (Right) Velocity-weighted APF.}
    \label{fig:goal_distance_comparison}
\end{figure}

\begin{itemize}
    \item Both methods successfully drive the distance to the goal toward zero (Figure~\ref{fig:goal_distance_comparison}). The velocity-weighted APF demonstrates a highly consistent rate of distance reduction, suggesting that by efficiently resolving local minima and preventing oscillations, it achieves a more reliable overall path efficiency.
\end{itemize}

\begin{figure}[h!]
    \centering
    \begin{subfigure}[b]{0.49\textwidth}
        \centering
        \includegraphics[width=\textwidth,height=5cm]{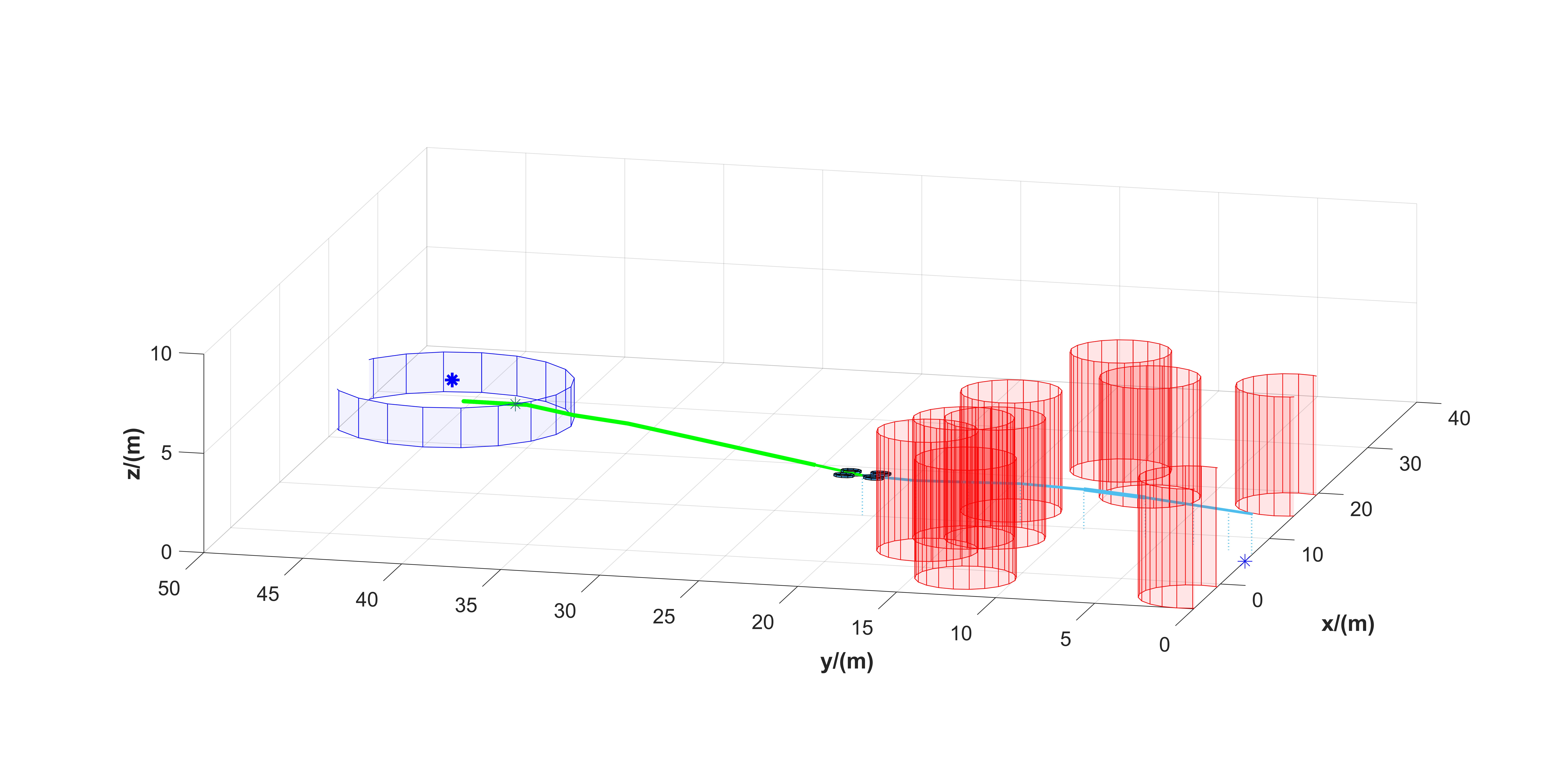}
        \caption{3D trajectory (View 2): alternative perspective emphasizing flight stability and smooth transitions.}
        \label{fig:3d_vel_v2}
    \end{subfigure}
    \hfill
    \begin{subfigure}[b]{0.49\textwidth}
        \centering
        \includegraphics[width=\textwidth,height=5cm]{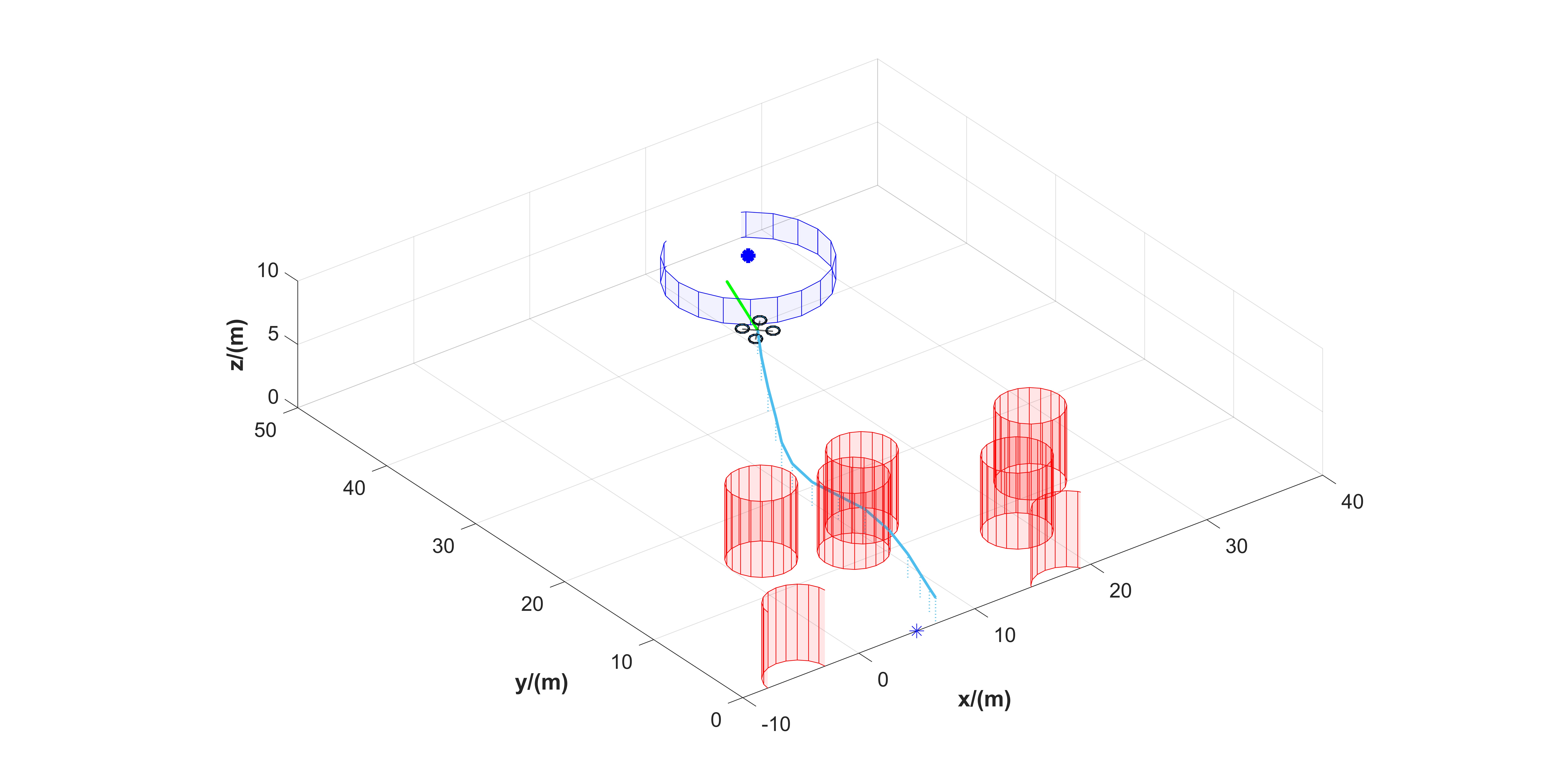}
        \caption{3D trajectory (View 3): close-up view of the path (light blue) smoothly converging towards the target region.}
        \label{fig:3d_vel_v3}
    \end{subfigure}
    \caption{3D visualization of the velocity-weighted APF trajectory navigating around cylindrical obstacles (red). The UAV successfully maintains a fixed altitude (z-axis) while weaving a stable, collision-free path toward the blue target ring.}
    \label{fig:3d_trajectories}
\end{figure}

Figures~\ref{fig:3d_vel_v2} and~\ref{fig:3d_vel_v3} provide multiple views of the velocity-weighted APF trajectory, demonstrating the algorithm's capability to generate stable paths in a cluttered 3D environment. These visualizations confirm the system's spatial awareness and its ability to achieve global movement while satisfying local constraints.

\section{Conclusion and Future Work}
\label{sec:conclusion}
\subsection{Conclusion}
This paper presented and validated a novel approach for enhancing UAV collision avoidance in dynamic and cluttered environments: the Direction and Relative Velocity Weighted Artificial Potential Field (APF) integrated with a Model Predictive Control (MPC) strategy. The methodology effectively addressed the critical limitations of conventional APF, particularly the notorious local minima issue and the lack of explicit dynamic obstacle threat assessment.

The simulation results confirmed the significant benefits of the proposed weighting function $\omega(\theta, v_e)$:

\begin{itemize}
    \item \textbf{Local minima resolution:} The velocity-weighted APF successfully navigated obstacle configurations specifically designed to induce local minima traps, demonstrating superior path integrity compared to the basic APF formulation (Figure~\ref{fig:static_minima}).
    \item \textbf{Enhanced safety and prediction:} The dynamic scaling of the repulsive force based on relative velocity resulted in smoother, more predictive avoidance manoeuvres, enhancing the safety margin during critical obstacle encounters (Figure~\ref{fig:obs_distance_comparison}).
    \item \textbf{Path efficiency:} By preventing stalling and excessive oscillation, the method maintained a highly consistent and efficient reduction in distance to the goal (Figure~\ref{fig:goal_distance_comparison}).
\end{itemize}

In conclusion, the integration of velocity and directional awareness into the APF framework provides a reliable and computationally efficient solution for autonomous path planning, making UAV operation safer and more feasible in complex, dynamic airspace.
\subsection{Future Work}

Future research efforts will focus on extending and refining the system in several key areas:

\begin{itemize}
    \item \textbf{Real-Time Implementation:} The methodology will be validated on a physical quadrotor platform to assess computational overhead and latency challenges in real-time operation, ensuring the approach remains practical for on-board processing.
    \item \textbf{Uncertainty and Sensing:} Incorporating sensor noise and measurement uncertainty models into the MPC objective function will enhance robustness against realistic sensing errors in position and velocity estimation.
    \item \textbf{Multi-Agent Coordination:} Extending the Velocity-Weighted APF to a multi-UAV system for coordinated navigation and collision avoidance, where the repulsive function is adapted for cooperative dynamic targets.
    \item \textbf{Energy Optimization:} While the current focus was on safety and stability, future work will involve modifying the MPC cost function ($\mathbf{R}$ matrix) to balance acceleration demands with energy consumption, aiming to smooth control inputs without compromising safety performance.
\end{itemize}

\bibliographystyle{IEEEtran}
\bibliography{cite}

\end{document}